\documentclass[a4paper,11pt]{article}
\usepackage{bm,mathrsfs,amsmath,amsthm,amssymb,amscd,longtable,array,graphicx}
\newcommand{\nc}{\newcommand}
\nc{\rnc}{\renewcommand}
\nc{\nn}{\nonumber}
\nc{\der}{{\partial}}
\rnc{\Im}{{\rm{Im}\,}}
\rnc{\Re}{{\rm{Re}\,}}
\nc{\db}{\displaybreak[0]\\}
\nc{\bra}{\langle}
\nc{\ket}{\rangle}
\nc{\bs}{\boldsymbol}

\newtheorem{theorem}{Theorem}[section]

\newtheorem{proposition}[theorem]{Proposition}

\theoremstyle{definition}
\newtheorem{definition}[theorem]{Definition}

\numberwithin{equation}{section}

\numberwithin{equation}{section}

\begin{document}%
%%%%%%%%%%%%%%%%%%%%%%%%%%%%%%%%%%%%%%%%%%%%%%%%%%%%%%%%%
%TITLE
%%%%%%%%%%%%%%%%%%%%%%%%%%%%%%%%%%%%%%%%%%%%%%%%%%%%%%%%%
%
\title{Izergin-Korepin analysis on
the projected wavefunctions of the generalized
free-fermion model}

\author{
Kohei Motegi \thanks{E-mail: kmoteg0@kaiyodai.ac.jp}
\\\\
{\it Faculty of Marine Technology, Tokyo University of Marine Science and Technology,}\\
 {\it Etchujima 2-1-6, Koto-Ku, Tokyo, 135-8533, Japan} \\
\\\\
\\
}

\date{\today}

\maketitle

\begin{abstract}
We apply the Izergin-Korepin analysis to the study of
the projected wavefunctions of the generalized free-fermion model.
We introduce a generalization of the $L$-operator
of the six-vertex model by Bump-Brubaker-Friedberg
and Bump-McNamara-Nakasuji.
We make the Izergin-Korepin analysis to characterize
the projected wavefunctions and show that they
can be expressed as a product of factors and certain symmetric functions
which generalizes the factorial Schur functions.
This result can be seen as a generalization
of the Tokuyama formula for the factorial Schur functions.
\end{abstract}

\section{Introduction}
Integrable lattice models \cite{Bethe,FST,Baxter,KBI}
are special classes of models in statistical physics
which many exact calculations are believed to be able to be done.
The most local object in integrable models is called as the $R$-matrix,
and its mathematical structure was revealed
in the mid 1980s \cite{Dr,J}. The underlying mathematical structure
was named as the quantum groups,
and the investigation of the quantum groups naturally lead to
immediate constructions of various $R$-matrices.

From the point of view of statistical physics,
$R$-matrices are the most local objects,
and the study on the $R$-matrices is a starting point.
The most important objects in statistical physics are partition functions.
For the case of integrable models, partition functions are objects
constructed from multiple $R$-matrices and
are determined by boundary conditions.
One of the most famous partition functions
in integrable lattice models are the domain wall
boundary partition functions which was first introduced and analyzed
in \cite{Ko,Iz}.
In recent years, a more general class of partition functions
which we shall call as the projected wavefunctions
are attracting attention in its relation
with algebraic combinatorics.
The projected wavefunctions are the projection of the 
off-shell Bethe vector of integrable models into a class of some simple states
labelled by the sequences of the particles or down spins.
For the case of the free-fermion model in an external field,
it was first shown by Bump-Brubaker-Friedberg \cite{BBF}
that the projected wavefunctions give a natural realization of the
Tokuyama combinatorial formula for the Schur functions \cite{To},
which is a one-parameter deformation of the Weyl character formula
(note there are pioneering works using the free-fermion model
implicitly in \cite{OkTo,HK1,HK2},
and the Izergin-Korepin analysis and observation of
the factorization phenomena on
the domain wall boundary partition functions
of the related models called as the Perk-Schultz (supersymmetric vertex)
model \cite{PS} and the Felderhof free-fermion model
\cite{Felderhof} in \cite{ZZ,FCWZ}.
There is also an application to the correlation functions in \cite{ZYZ}).
This observation triggered studies on finding
various generalizations and variations of the
Tokuyama-type formula for symmetric functions
\cite{Iv,BBCG,Tabony,BMN,HK,BS,BBB,LMP,dualsymplectic}
such as the factorial Schur functions and symplectic Schur functions,
and an interesting notion was introduced furthermore
which the number theorists call as the metaplectic ice.

In this paper, we analyze the free-fermion model
using the method initiated by Izergin-Korepin \cite{Ko,Iz}.
The method was developed by them in order to find the explicit expression
of polynomials representing the domain wall boundary partition functions
of the $U_q(sl_2)$ six-vertex model,
from which the famous Izergin-Korepin determinant formula was found.
The Izergin-Korepin analysis is the important method
to study variants of the domain wall boundary partition functions.
For example, it was applied to the domain wall boundary partition functions
of the $U_q(sl_2)$ six-vertex model with reflecting end
by Tsuchiya \cite{Tsuchiya} to find its determinant formula.
Extending the Izergin-Korepin analysis
to more general class of partition functions
are also important.
Wheeler \cite{Wheeler}
invented a method to extend the Izergin-Korepin analysis
on a class of partition functions called the scalar products.
And in our very recent work \cite{Motr}, we extended the
Izergin-Korepin analysis to study the projected wavefunctions
of the $U_q(sl_2)$ six-vertex model.
The resulting symmetric polynomials representing the projected wavefunctions
contains the Grothendieck polynomials as a special case
when the six-vertex model reduces to the five-vertex model \cite{MS,Motegi,WZ}.
We apply this technique to study the free-fermion model in an external field.
To this end, we first introduce an ultimate generalization of
the $L$-operator by introducing the inhomogeneous parameters
and factorial parameters. We use an inhomogeneous version of
the generalized $L$-operator in our forthcoming paper \cite{MSW}
having two types of factorial parameters, which
generalizes the factorial $L$-operator by Bump-McNamara-Nakasuji \cite{BMN}.
We next view the projected wavefunctions as a function of the inhomogeneous parameters
and characterize its properties by using the Izergin-Korepin analysis.
We then show that the product of
factors and certain symmetric functions
satisfies all the required properties the projected wavefunctions must satisfy.
The result is a generalization of the \cite{BBF} and \cite{BMN},
hence can be viewed as a generalization of the
Tokuyama for the factorial Schur functions.
The Izergin-Korepin analysis views the partition functions
as functions of inhomogeneous parameters in the quantum spaces,
whereas the arguments initiated in \cite{BBF}
view the partition functions as functions of
the free parameter in the auxiliary spaces.
The comparison of the two different ways of arguments seems to be interesting.

We will use the results of the projected wavefunctions to the
algebraic combinatorial study of the generalized Schur functions
\cite{MSW}.
For example, two ways of evaluations of the same partition functions
can lead to integrable model constructions of algebraic identities
of the symmetric functions.
For example, two ways of evaluations
of the domain wall boundary partition functions,
a direct evaluation and an indirect evaluation using
the completeness relation and the projected wavefunctions,
can give rise to the dual Cauchy formula of the generalized Schur functions.
This idea can also be applied to partition functions of integrable models
under reflecting boundary to give dual Cauchy identities of the
generalized symplectic Schur functions.
Further detailed Izergin-Korepin analysis
on the domain wall boundary partition functions
and the dual projected wavefunctions are required for the studies.

There are also studies on deriving Cauchy identities
using the domain wall boundary partition functions like an intertwiner,
invented in \cite{WZnew}.
Deriving algebraic combinatorial properties of symmetric functions using
their integrable model realizations is an active line of research.
See \cite{BW,BWZ,KS,Korff,Borodin} for more examples on Cauchy-type identities
and more recent studies on the
Littlewood-Richardson coefficients by \cite{WZ,WZLRone}.

In any case, in order to do these studies, we first of all have to
find out what are the explicit functions
representing the projected wavefunctions.
We think the Izergin-Korepin analysis presented in this paper
is a fairly simple way to find out the explicit forms.

This paper is organized as follows.
In the next section, we first list the generalized $L$-operator
and introduce the projected wavefunctions.
In section 3, we make the Izergin-Korepin analysis
and list the properties needed to determine the explicit form
of the projected wavefunctions.
In section 4, we show that the product of factors
and certain symmetric functions
satisfies all the required properties extracted from the
Izergin-Korepin analysis, which means that the product
is the explicit form of the projected wavefunctions.
Section 5 is devoted to the conclusion of this paper.

\section{The generalized free-fermion model and the projected wavefunctions}
The most fundamental objects in integrable lattice models
are the $R$-matrices and $L$-operators.
The $R$-matrix of the free-fermion model we treat in this paper is given by
\begin{eqnarray}
R_{ab}(z)=\left( 
\begin{array}{cccc}
1+tz & 0 & 0 & 0 \\
0 & t(1-z) & t+1 & 0 \\
0 & (t+1)z & z-1 & 0 \\
0 & 0 & 0 & z+t
\end{array}
\right), \label{rmatrix}
\end{eqnarray}
acting on the tensor product $W_a \otimes W_b$
of the complex two-dimensional space $W_a$.

\begin{figure}[ht]
\includegraphics[width=15cm]{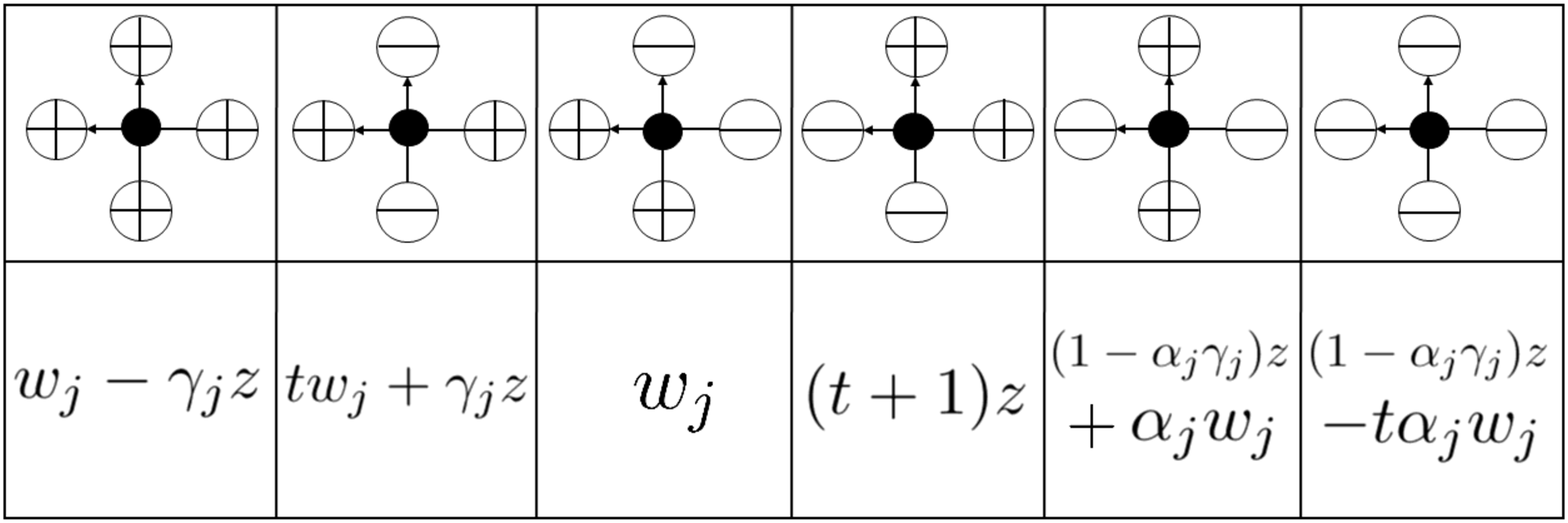}
\caption{The $L$-operator $L_{aj}(z,w_j,\alpha_j,\gamma_j)$
\eqref{generalizedloperator}.
The horizontal line is the space $W_a$, and the vertical line
is the space $\mathcal{F}_j$.
}
\label{pictureloperator}
\end{figure}

The $L$-operator of the free-fermion model
we use as bulk pieces of the projected wavefunctions in this paper is given by
\begin{align}
&L_{aj}(z,w_j,\alpha_j,\gamma_j) \nonumber \\
=&\left( 
\begin{array}{cccc}
w_j-\gamma_j z & 0 & 0 & 0 \\
0 & tw_j+\gamma_j z & w_j & 0 \\
0 & (t+1)z & \alpha_j w_j+(1-\alpha_j \gamma_j)z & 0 \\
0 & 0 & 0 & -t \alpha_j w_j+(1-\alpha_j \gamma_j)z
\end{array}
\right), \label{generalizedloperator}
\end{align}
acting on the tensor product $W_a \otimes \mathcal{F}_j$ of the space
$W_a$ and the two-dimensional Fock space at the
$j$th site $\mathcal{F}_j$.

The parameters $w_j$, $\alpha_j$ and $\gamma_j$ can be
regarded as parameters associated with the quantum space $\mathcal{F}_j$.
The $L$-operators giving the
Schur functions \cite{BBF} and
factorial Schur functions \cite{BMN}
is a special limit of the generalized $L$-operator
\eqref{generalizedloperator}
given by
\begin{eqnarray}
L_{aj}(z,1,0,0)=\left( 
\begin{array}{cccc}
1 & 0 & 0 & 0 \\
0 & t & 1 & 0 \\
0 & (t+1)z & z & 0 \\
0 & 0 & 0 & z
\end{array}
\right),
\end{eqnarray}
\begin{eqnarray}
L_{aj}(z,1,\alpha_j,0)=\left( 
\begin{array}{cccc}
1 & 0 & 0 & 0 \\
0 & t & 1 & 0 \\
0 & (t+1)z & \alpha_j+z & 0 \\
0 & 0 & 0 & -t \alpha_j+z
\end{array}
\right),
\end{eqnarray}
respectively.

The $L$-operator \eqref{generalizedloperator}
together with the $R$-matrix \eqref{rmatrix}
satisfies the $RLL$ relation
\begin{align}
&R_{ab}(z_1/z_2)L_{aj}(z_1,w_j,\alpha_j,\gamma_j)
L_{bj}(z_2,w_j,\alpha_j,\gamma_j) \nonumber \\
=&L_{bj}(z_2,w_j,\alpha_j,\gamma_j)L_{aj}(z_1,w_j,\alpha_j,\gamma_j)
R_{ab}(z_1/z_2), \label{RLL}
\end{align}
acting on $W_a \otimes W_b \otimes \mathcal{F}_j$.

Let us denote the orthonormal basis of $W_a$ and its dual as
$\{|0 \rangle_a, |1 \rangle_a \}$ and $\{{}_a \langle 0|, {}_a \langle 1|\}$,
and the orthonormal basis of $\mathcal{F}_j$ and its dual as
$\{|0 \rangle_j, |1 \rangle_j \}$ and $\{{}_j \langle 0|, {}_j \langle 1|\}$.
The matrix elements of the $L$-operator can be written as
$
{}_a \langle \gamma| {}_j \langle \delta | L_{a j}(z,w_j,\alpha_j,\gamma_j)
|\alpha \rangle_a | \beta \rangle_j
$, which we will use this form in the next section.
See Figure \ref{pictureloperator} for a pictorial description
of the $L$-operator \eqref{generalizedloperator}.

The $R$-matrices and
the $L$-operators have origins in statistical physics,
and $| 0 \rangle$ or its dual $\langle 0|$
can be regarded as a hole state,
while $| 1 \rangle$ or its dual $\langle 1|$
can be interpretted as a particle state
from the point of view of statistical physics.
We sometimes use the terms hole states and particle states
to describe states constructed from
$| 0 \rangle$, $\langle 0|$, $| 1 \rangle$ and $\langle 1|$
since they are convenient for the description of the states.
In the quantum inverse scattering method,
the Fock spaces $W_a$ and $\mathcal{F}_j$
are usually called the auxiliary and quantum spaces, respectively.

For later convenience,
we also define the following Pauli spin operators
$\sigma^+$ and $\sigma^-$ as operators acting on the (dual) orthonomal
basis as
\begin{align}
&\sigma^+|1 \rangle=|0 \rangle, \ 
\sigma^+|0 \rangle=0, \ 
\langle 0|\sigma^+=\langle 1|, \
\langle 1|\sigma^+=0, 
\\
&\sigma^-|0 \rangle=|1 \rangle, \
\sigma^-|1 \rangle=0, \
\langle 1|\sigma^-=\langle 0|, \
\langle 0|\sigma^-=0.
\end{align}

\begin{figure}[ht]
\includegraphics[width=12cm]{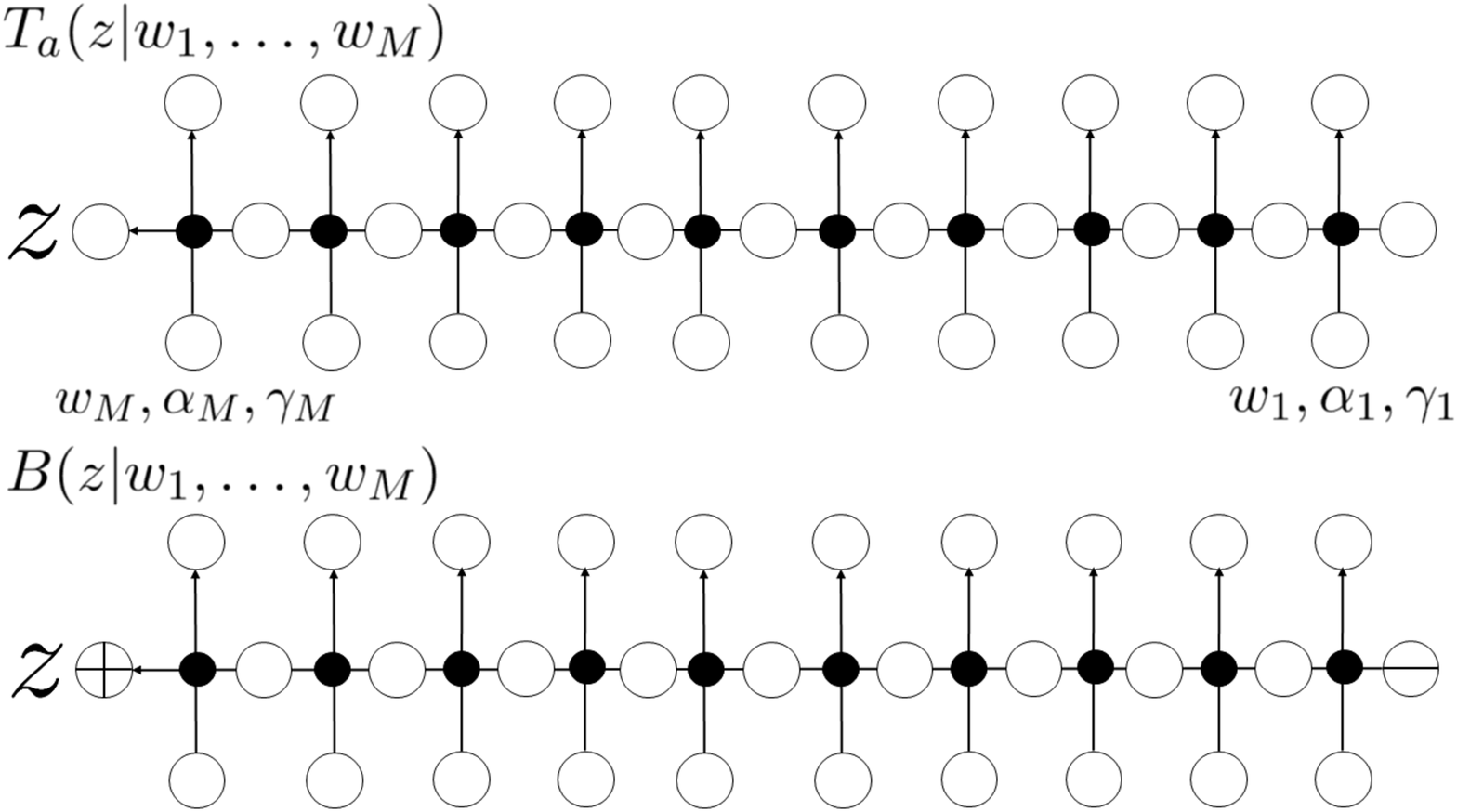}
\caption{The monodromy matrix $T_a(z|w_1,\dots,w_M)$
\eqref{monodromy} (top)
and the $B$-operator $B(z|w_1,\dots,w_M)$ (bottom).}
\label{picturemonodromy}
\end{figure}

\begin{figure}[ht]
\includegraphics[width=12cm]{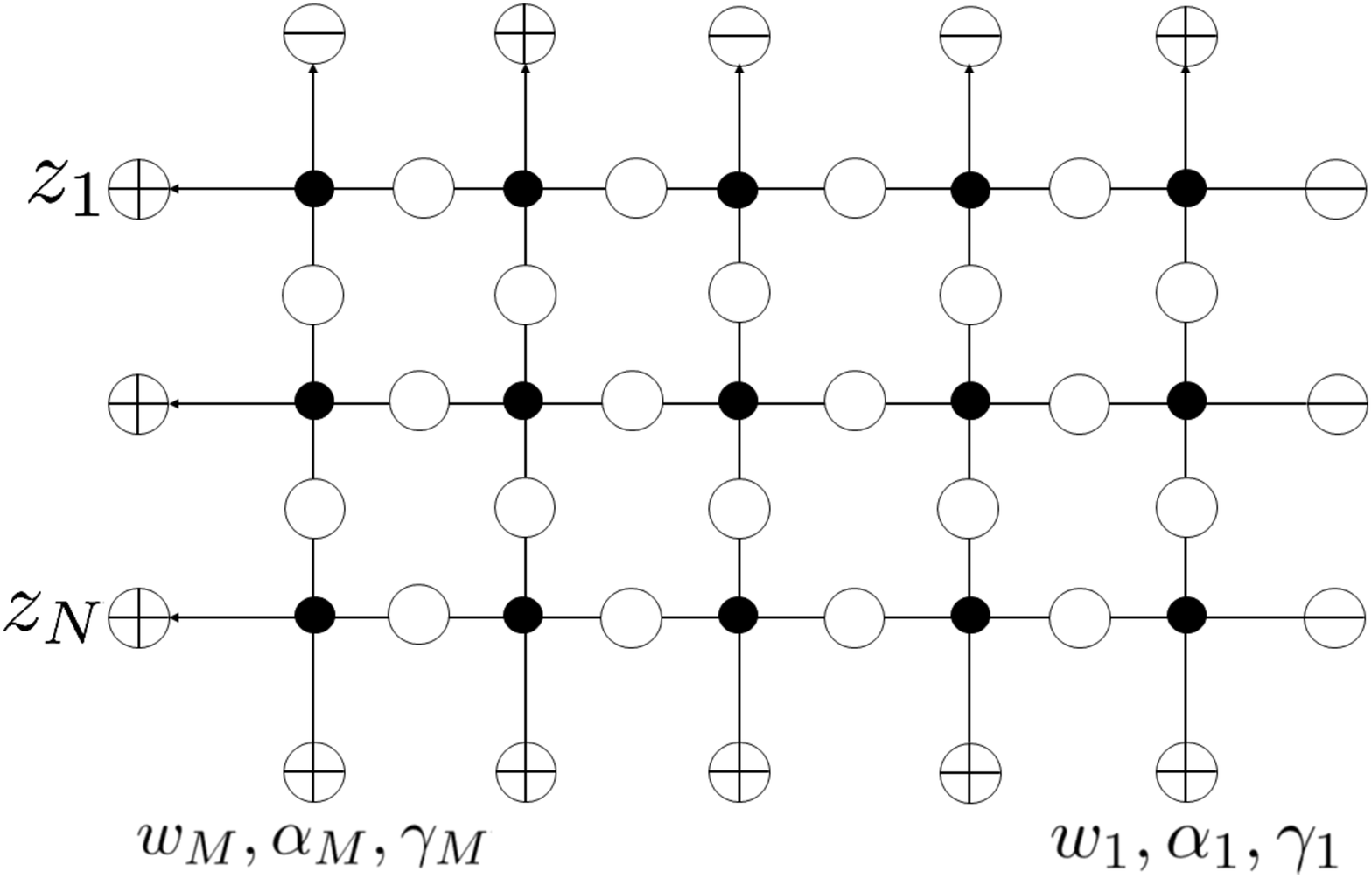}
\caption{The projected wavefunctions
$W_{M,N}(z_1,\dots,z_N|w_1,\dots,w_M|x_1,\dots,x_N)$
\eqref{ordinaryDWBPF}.
This figure illustrates the case
$M=5$, $N=3$, $x_1=2$, $x_2=3$, $x_3=5$.
}
\label{ordinarypictureDWBPF}
\end{figure}

To construct projected wavefunctions,
we introduce
the monodromy matrix $T_a(z|w_1,\dots,w_M)$
(Figure \ref{picturemonodromy} top) from the generalized
$L$-operator \eqref{generalizedloperator} as
\begin{align}
T_{a}(z|w_1,\dots,w_M)&=L_{a M}(z,w_M,\alpha_M,\gamma_M) \cdots
L_{a 1}(z,w_1,\alpha_1,\gamma_1)
\nonumber \\
&=
\begin{pmatrix}
A(z|w_1,\dots,w_M) & B(z|w_1,\dots,w_M)  \\
C(z|w_1,\dots,w_M) & D(z|w_1,\dots,w_M)
\end{pmatrix}_{a} \in \mathrm{End}(W_a \otimes \mathcal{F}_1 \otimes
\cdots \otimes \mathcal{F}_M).
\label{monodromy}
\end{align}

The matrix elements
$
A(z|w_1,\dots,w_M),
B(z|w_1,\dots,w_M),
C(z|w_1,\dots,w_M)$ \\
and $D(z|w_1,\dots,w_M)$
are called as the $ABCD$ operators, which are $2^M \times 2^M$ matrices
acting on the tensor product of the quantum spaces
$\mathcal{F}_1\otimes \dots \otimes \mathcal{F}_M$.

To create projected wavefunctions,
what is important is the $B$-operator $B(z|w_1,\dots,w_M)$
(Figure \ref{picturemonodromy} bottom)
which has the role of creating particles in the quantum spaces
$\mathcal{F}_1 \otimes \cdots \otimes \mathcal{F}_M$.
We next introduce the following state vector \\
$|\Phi_{M,N}(z_1,\dots,z_N|w_1,\dots,w_M) \rangle
\in \mathcal{F}_1 \otimes \cdots \otimes \mathcal{F}_M$
using the $B$-operators as
\begin{align}
|\Phi_{M,N}(z_1,\dots,z_N|w_1,\dots,w_M)
\rangle=B(z_1|w_1,\dots,w_M) \cdots B(z_N|w_1,\dots,w_M)|\Omega \rangle_M,
\label{ordinarystatevector}
\end{align}
where $|\Omega \rangle_M:=|0 \rangle_1 \otimes \cdots \otimes |0 \rangle_M
\in \mathcal{F}_1 \otimes \cdots \otimes \mathcal{F}_M$
is the vacuum state in the tensor product of quantum spaces.

Due to the so-called ice rule of the $L$-operator
${}_{a} \langle \gamma | {}_{j} \langle \delta |
L_{aj}(z,w_j,\alpha_j,\gamma_j)|\alpha \rangle_a |\beta \rangle_j=0$
unless $\alpha+\beta=\gamma+\delta$, each $B$-operator creates one particle
in the quantum spaces. This fact and that the state vector
\eqref{ordinarystatevector} is constructed from $N$-layers
of the $B$-operators acting on the vacuum state
$|\Omega \rangle_M$, the state vector \eqref{ordinarystatevector}
is an $N$-particle state for $N \le M$.
To construct a nonvanishing inner product,
we introduce the dual $N$-particle state
\begin{align}
\langle x_1 \cdots x_N|
&=(_1 \langle 0| \otimes \cdots \otimes {}_M \langle 0|)
\prod_{j=1}^N \sigma^+_{x_j}
\in \mathcal{F}_1^* \otimes \cdots \otimes \mathcal{F}_M^*
, \label{ordinarydualparticleconfiguration}
\end{align}
which are states labelling the configurations
of particles
$1 \le x_1 < x_2 < \cdots < x_N \le M$.

The projected wavefunctions
$W_{M,N}(z_1,\dots,z_N|w_1,\dots,w_M|x_1,\dots,x_N)$
is defined as the inner product between the state vector
(off-shell Bethe vector)
$|\Phi_{M,N}(z_1,\dots,z_N|w_1,\dots,w_M) \rangle$
and the $N$-particle state $\langle x_1 \cdots x_N|$
\begin{align}
W_{M,N}(z_1,\dots,z_N|w_1,\dots,w_M|x_1,\dots,x_N)
=\langle x_1 \cdots x_N|\Phi_{M,N}(z_1,\dots,z_N|w_1,\dots,w_M) \rangle.
\label{ordinaryDWBPF}
\end{align}
See Figure \ref{ordinarypictureDWBPF} for a pictorial description
of \eqref{ordinaryDWBPF}.

In the next section, we examine the properties of the 
projected wavefunctions.
Here we just remark that
the projected wavefunctions of the free-fermion model treated in this paper
is not symmetric with respect to the spectral parameters
$\{ z_1,\dots,z_N \}$.
This is in contrast to the case of the projected wavefunctions 
of the $U_q(sl_2)$ six-vertex model,
where they are symmetric with respect to the
spectral variables, in which case the Grothendieck polynomials
and its quantum group deformation appears.
This fact for the properties of the spectral variables
of the free-fermion model
lead to the Tokuyama formula \cite{To} for the Schur functions,
as was first found in \cite{BBF}.

\section{Izergin-Korepin analysis}

By the Izergin-Korepin analysis,
we examine the properties
of the projected wavefunctions \\
$W_{M,N}(z_1,\dots,z_N|w_1,\dots,w_M|x_1,\dots,x_N)$
in this section.

\begin{proposition} 
\label{ordinarypropertiesfordomainwallboundarypartitionfunction}
The projected wavefunctions
$W_{M,N}(z_1,\dots,z_N|w_1,\dots,w_M|x_1,\dots,x_N)$
satisfies the following properties. \\
\\
 (1) $W_{M,N}(z_1,\dots,z_N|w_1,\dots,w_M|x_1,\dots,x_N)$
is a polynomial of degree $N$ in $w_M$.
\\
 (2) The projected wavefunctions $W_{M,N}(z_{\sigma(1)},\dots,z_{\sigma(N)}|w_1,\dots,w_M|x_1,\dots,x_N)$ with the ordering of the spectral parameters permuted
$z_{\sigma(1)}, \dots, z_{\sigma(N)}$, $\sigma \in S_N$ are related with
the unpermuted one
$W_{M,N}(z_1,\dots,z_N|w_1,\dots,w_M|x_1,\dots,x_N)$
by the following relation
\begin{align}
&\prod_{\substack{1 \le j < k \le N \\ \sigma(j) > \sigma(k)}}
(z_{\sigma(j)}+t z_{\sigma(k)})
W_{M,N}(z_1,\dots,z_N|w_1,\dots,w_M|x_1,\dots,x_N) \nonumber \\
=&
\prod_{\substack{1 \le j < k \le N \\ \sigma(j) > \sigma(k)}}
(z_{\sigma(k)}+t z_{\sigma(j)})
W_{M,N}(z_{\sigma(1)},\dots,z_{\sigma(N)}|w_1,\dots,w_M|x_1,\dots,x_N).
\label{permutationwavefunction}
\end{align}
\\
(3) The following recursive relations between the
projected wavefunctions hold if $x_N=M$:
\begin{align}
&W_{M,N}(z_1,\dots,z_N|w_1,\dots,w_M|x_1,\dots,x_N)
|_{w_M=\gamma_M z_N}
\nonumber \\
=&\gamma_M^N z_N \prod_{j=1}^{N-1} (z_j+t z_N)
\prod_{j=1}^{M-1} \{ (1-\alpha_j \gamma_j)z_N+\alpha_j w_j \}
\nonumber \\
&\times W_{M-1,N-1}(z_1,\dots,z_{N-1}|w_1,\dots,w_{M-1}|x_1,\dots,x_{N-1})
. \label{ordinaryrecursionwavefunction}
\end{align}
When evaluated at $w_M=0$, we have
\begin{align}
W_{M,N}(z_1,\dots,z_N|w_1,\dots,w_M|x_1,\dots,x_N)
|_{w_M=0}=0. \label{ordinaryrecursionwavefunctionaddition}
\end{align}

If $x_N \neq M$, the following factorizations hold for the
projected wavefunctions:
\begin{align}
&W_{M,N}(z_1,\dots,z_N|w_1,\dots,w_M|x_1,\dots,x_N)
 \nonumber \\
=&\prod_{j=1}^N (w_M-\gamma_M z_j)
W_{M-1,N}(z_1,\dots,z_N|w_1,\dots,w_{M-1}|x_1,\dots,x_N).
\label{ordinaryrecursionwavefunction2}
\end{align}
\\
(4) The following holds for the case $N=1$, $x_N=M$
\begin{align}
&
W_{M,1}(z|w_1,\dots,w_M|M)=w_M \prod_{k=1}^{M-1} \{ (1-\alpha_k \gamma_k)z
+\alpha_k w_k \}.
\label{ordinaryinitialrecursion}
\end{align}
\end{proposition}

\begin{proof}
Let us first show Properties (1) and (3)
for the case $x_N=M$.

To show Property (1) when $x_N=M$,
we first express the projected wavefunctions
in terms of the vertical transfer matrix
\begin{align}
\mathcal{T}_j^N(w_j;z_1,\dots,z_N)
=&L_{a_1 j}(z_1,w_j,\alpha_j,\gamma_j) \cdots
L_{a_N j}(z_N,w_j,\alpha_j,\gamma_j) \nonumber \\
&
\in \mathrm{End}(W_{a_1} \otimes \cdots \otimes W_{a_N} \otimes \mathcal{F}_j).
\end{align}
Using this vertical transfer matrix, the projected wavefunctions
can be rewritten as
\begin{align}
&W_{M,N}(z_1,\dots,z_N|w_1,\dots,w_M|x_1,\dots,x_{N-1},M) \nonumber \\
=&\langle 0|^{\otimes N}
{}_{M} \langle 1| \langle x_1 \cdots x_{N-1}|
\mathcal{T}_{M}^N(w_M;z_1,\dots,z_N) \cdots
\mathcal{T}_{1}^N(w_1;z_1,\dots,z_N)
|1 \rangle^{\otimes N}
|\Omega \rangle_M, \\
&\langle 0|^{\otimes N}=
 {}_{a_1} \langle 0| \otimes \cdots \otimes {}_{a_N} \langle 0|, \\
&|1 \rangle^{\otimes N}=
 |1 \rangle_{a_1} \otimes \cdots \otimes |1 \rangle_{a_N}.
\label{anotherexpression}
\end{align}

Inserting the completeness relation
in one particle sector
\begin{align}
&\sum_{j=1}^{N}
|0^{j-1},1,0^{N-j} \rangle \langle 0^{j-1},1,0^{N-j} |=\mathrm{Id}, \\
&|0^{j-1},1,0^{N-j} \rangle=|0 \rangle_{a_1} \otimes \cdots \otimes
|0 \rangle_{a_{j-1}} \otimes |1 \rangle_{a_j} \otimes
|0 \rangle_{a_{j+1}} \otimes \cdots \otimes |0 \rangle_{a_N}, \\
&\langle 0^{j-1},1,0^{N-j} |={}_{a_1} \langle 0 | \otimes \cdots \otimes
{}_{a_{j-1}} \langle 0 | \otimes {}_{a_j} \langle 1 | \otimes
{}_{a_{j+1}} \langle 0 | \otimes \cdots \otimes {}_{a_N} \langle 0 |,
\end{align}
into \eqref{anotherexpression}, we have
\begin{align}
&W_{M,N}(z_1,\dots,z_N|w_1,\dots,w_M|x_1,\dots,x_{N-1},M) \nonumber \\
=&\sum_{j=1}^{N} \langle 0|^{\otimes N} {}_{M} \langle 1|
\mathcal{T}_M^N(w_M;z_1,\cdots,z_N)
|0^{j-1},1,0^{N-j} \rangle |0 \rangle_M \nonumber \\
&\times \langle x_1 \cdots x_{N-1}|
\langle 0^{j-1},1,0^{N-j} | \mathcal{T}_{M-1}^N(w_{M-1};z_1,\cdots,z_N)
\cdots \mathcal{T}_1^N(w_1;z_1,\cdots,z_N)
|\Omega \rangle_{M-1} |1 \rangle^{\otimes N}.
\label{anotherexpressionexpansion}
\end{align}
In the right hand side of \eqref{anotherexpressionexpansion},
the parameter $w_M$ depends only on \\
$\langle 0|^{\otimes N} {}_{M} \langle 1|
\mathcal{T}_M^N(w_M;z_1,\cdots,z_N)
|0^{j-1},1,0^{N-j} \rangle |0 \rangle_M
$, whose matrix elements can be easily calculated
from its graphical representation as
\begin{align}
\langle 0|^{\otimes N} {}_{M} \langle 1|
\mathcal{T}_M^N(w_M;z_1,\cdots,z_N)
|0^{j-1},1,0^{N-j} \rangle |0 \rangle_M
=w_M \prod_{k=1}^{j-1}(tw_M+\gamma_M z_k)
\prod_{k=j+1}^N (w_M-\gamma_M z_k). \label{matrixelementuse}
\end{align}
Since the matrix elements \eqref{matrixelementuse}
is a polynomial of degree $N$ in $w_M$,
one finds that the projected wavefunctions is a polynomial
of degree $N$ in $w_M$.

Let us next show Property (3) for the case $x_N = M$.
We first remark that since the projected wavefunctions
$W_{M,N}(z_1,\dots,z_N|w_1,\dots,w_M|x_1,\dots,x_N)$
is a polynomial of degree $N$ in $w_M$,
one needs to evaluate $N+1$ distinct points in $w_M$
for the Izergin-Korepin trick to be successful.
\eqref{ordinaryrecursionwavefunction} is the result of the
evaluation at the point $w_M=\gamma_M z_N$.
The $(N-1)$ points $w_M=\gamma_M z_j$, $j=1,\dots,N-1$
can be evaluated using Property (2),
hence if one shows that certain functions satisfy Property (2),
it remains to consider the evaluation at $w_M=\gamma_M z_N$.
The evaluation at $w_M=\gamma_M z_N$ essentially gives evaluations
at $N$ distinct points.
We need one more point to be evaluated.
An easy point to be evaluated is $w_M=0$, whose result is
\eqref{ordinaryrecursionwavefunctionaddition}.
Let us show these two results of the evaluations.

The recursion relation
\eqref{ordinaryrecursionwavefunction}
can be shown as follows.
First, from the decomposition
\eqref{anotherexpressionexpansion}
and the explicit form of the matrix elements \eqref{matrixelementuse},
one finds that after the substitution $w_M=\gamma_M z_N$,
only the term $j=N$ of the sum in \eqref{anotherexpressionexpansion}
survives and we have
\begin{align}
&W_{M,N}(z_1,\dots,z_N|w_1,\dots,w_M|x_1,\dots,x_{N-1},M)|_{w_M=\gamma_M z_N}
\nonumber \\
=&\gamma_M^N z_N \prod_{j=1}^{N-1}(z_j+tz_N) \nonumber \\
&\times \langle x_1 \cdots x_{N-1}|
\langle 0^{N-1},1 | \mathcal{T}_{M-1}^N(w_{M-1};z_1,\cdots,z_N)
\cdots \mathcal{T}_1^N(w_1;z_1,\cdots,z_N)
|\Omega \rangle_{M-1} |1 \rangle^{\otimes N}.
\label{anotherexpressionexpansiontowardsfinal}
\end{align}
Since we can calculate the right hand side of
\eqref{anotherexpressionexpansiontowardsfinal} furthermore as
\begin{align}
&\langle x_1 \cdots x_{N-1}|
\langle 0^{N-1},1 | \mathcal{T}_{M-1}^N(w_{M-1};z_1,\cdots,z_N)
\cdots \mathcal{T}_1^N(w_1;z_1,\cdots,z_N)
|\Omega \rangle_{M-1} |1 \rangle^{\otimes N}
\nonumber \\
=&\langle x_1 \cdots x_{N-1}|
\langle 0^{N-1}| \mathcal{T}_{M-1}^{N-1}(w_{M-1};z_1,\cdots,z_{N-1})
\cdots \mathcal{T}_1^{N-1}(w_1;z_1,\cdots,z_{N-1})
|\Omega \rangle_{M-1} |1 \rangle^{\otimes N-1} \nonumber \\
&\times
{}_{a_N} \langle 1| {}_{M-1} \langle \Omega|
L_{a_N,M-1}(z_N,w_{M-1},\alpha_{M-1},\gamma_{M-1})
\cdots
L_{a_N,1}(z_N,w_1,\alpha_1,\gamma_1)
|1 \rangle_{a_N} |\Omega \rangle_{M-1} \nonumber \\
=&W_{M-1,N-1}(z_1,\dots,z_{N-1}|w_1,\dots,w_{M-1}|x_1,\dots,x_{N-1})
\prod_{j=1}^{M-1} \{ (1-\alpha_j \gamma_j)z_N+\alpha_j w_j \},
\end{align}
we can express the evaluation of
$W_{M,N}(z_1,\dots,z_N|w_1,\dots,w_M|x_1,\dots,x_N)$
at $w_M=\gamma_M z_N$ as
\begin{align}
&W_{M,N}(z_1,\dots,z_N|w_1,\dots,w_M|x_1,\dots,x_N)
|_{w_M=\gamma_M z_N}
\nonumber \\
=&\gamma_M^N z_N \prod_{j=1}^{N-1} (z_j+t z_N)
\prod_{j=1}^{M-1} \{ (1-\alpha_j \gamma_j)z_N+\alpha_j w_j \}
\nonumber \\
&\times W_{M-1,N-1}(z_1,\dots,z_{N-1}|w_1,\dots,w_{M-1}|x_1,\dots,x_{N-1}).
\end{align}

The evaluation at $w_M=0$
\eqref{ordinaryrecursionwavefunctionaddition}
can be easily seen by the expansion
\eqref{anotherexpressionexpansion}
and the fact that all the matrix elements \eqref{matrixelementuse}
contain the factor $w_M$.

Properties (1) and (3) for the case $x_N \neq M$ can be
shown much easier.
Using the ice rule
${}_{a} \langle \gamma | {}_{j} \langle \delta |
L_{aj}(z,w_j,\alpha_j,\gamma_j)|\alpha \rangle_a |\beta \rangle_j=0$,
one can easily finds the following factorization
\begin{align}
&W_{M,N}(z_1,\dots,z_N|w_1,\dots,w_M|x_1,\dots,x_{N}) \nonumber \\
=&\langle 0|^{\otimes N}
{}_{M} \langle 0| \langle x_1 \cdots x_{N}|
\mathcal{T}_{M}^N(w_M;z_1,\dots,z_N) \cdots
\mathcal{T}_{1}^N(w_1;z_1,\dots,z_N)
|1 \rangle^{\otimes N}
|\Omega \rangle_M
\nonumber \\
=&\langle 0|^{\otimes N}
{}_{M} \langle 0|
\mathcal{T}_{M}^N(w_{M};z_1,\dots,z_N)
|0 \rangle^{\otimes N}
|0 \rangle_M \nonumber \\
&\times \langle 0|^{\otimes N}
\langle x_1 \cdots x_{N}|
\mathcal{T}_{M-1}^N(w_{M-1};z_1,\dots,z_N) \cdots
\mathcal{T}_{1}^N(w_1;z_1,\dots,z_N)
|1 \rangle^{\otimes N}
|\Omega \rangle_{M-1} \nonumber \\
=&
\prod_{j=1}^N
{}_{a_j} \langle 0 | {}_{M} \langle 0 |
L_{a_j,M}(z_{j},w_M,\alpha_M,\gamma_M)|0 \rangle_{a_j} |0 \rangle_M
\nonumber \\
&\times W_{M-1,N}(z_1,\dots,z_N|w_1,\dots,w_{M-1}|x_1,\dots,x_N)
 \nonumber \\
=&\prod_{j=1}^N (w_M-\gamma_M z_j)
W_{M-1,N}(z_1,\dots,z_N|w_1,\dots,w_{M-1}|x_1,\dots,x_N).
\end{align}
This shows Properties (1) and (3) for the case $x_N \neq M$.

Property (2) can be shown as follows.
Using the $RLL$ relation repeatedly, one gets
the intertwining relation between the monodromy matrices
\begin{align}
&R_{ab}(z_1/z_2)T_{a}(z_1|w_1,\dots,w_M)T_{b}(z_2|w_1,\dots,w_M)
\nonumber \\
=&T_{b}(z_2|w_1,\dots,w_M)T_{a}(z_1|w_1,\dots,w_M)R_{ab}(z_1/z_2).
\label{RTT}
\end{align}
An element of the intertwining relation \eqref{RTT}
gives the commutation relation between the $B$-operators
\begin{align}
&(z_2+tz_1)B(z_1|w_1,\dots,w_M)B(z_2|w_1,\dots,w_M)
\nonumber \\
=&B(z_2|w_1,\dots,w_M)B(z_1|w_1,\dots,w_M)(z_1+tz_2).
\label{commutationBoperators}
\end{align}
Since the projected wavefunctions
$W_{M,N}(z_1,\dots,z_N|w_1,\dots,w_M|x_1,\dots,x_N)$
\eqref{ordinaryDWBPF} are constructed from $N$-layers
of $B$-operators, the effect of reordering the spectral parameters
of the $B$-operators can be traced using the
commutation relation \eqref{commutationBoperators}.

What finally remains is Property (4), which can also easily calculated.
\end{proof}

Before presenting the solution in the next section,
we explain here why the Izergin-Korepin analysis
uniquely defines the projected wavefunctions.
The idea is based on the following fact:
if there are two polynomials $f(w)$ and $g(w)$ of $w$
of degree $N$, and the evaluations of the two polynomials at
$N+1$ distinct points are the same ($f(w)=g(w)$
for $w=z_j$, $j=1,\dots,N+1$ such that $z_j \neq z_k$, $j \neq k$),
then the two polynomials are exactly the same, i.e., $f(w)=g(w)$ for all $w$.
The idea of Izergin-Korepin analysis is to relate
the projected
wavefunctions $W_{M,N}(z_1,\dots,z_N|w_1,\dots,w_M|x_1,\dots,x_N)$
with smaller ones by using the above fact.
The point is to regard
$W_{M,N}(z_1,\dots,z_N|w_1,\dots,w_M|x_1,\dots,x_N)$
as a polynomial of a single variable $w_M$.
By Property (1) in
Proposition \ref{ordinarypropertiesfordomainwallboundarypartitionfunction},
$W_{M,N}(z_1,\dots,z_N|w_1,\dots,w_M|x_1,\dots,x_N)$ is a polynomial
of degree $N$ in $w_M$.
If $x_N=M$, one can evaluate the projected wavefunction
at the following $N+1$ points
$w_M=\gamma_M z_j$, $j=1,\dots,N$, $w_M=0$.
The evaluations at $w_M=\gamma_M z_N$ and $w_M=0$
can be obtained by its graphical representation
and can be expressed by using the smaller projected wavefuntion
$W_{M-1,N-1}(z_1,\dots,z_{N-1}|w_1,\dots,w_{M-1}|x_1,\dots,x_{N-1})$,
which is \eqref{ordinaryrecursionwavefunction} and
\eqref{ordinaryrecursionwavefunctionaddition} of Property (3).
The evaluations at $(N-1)$ points $w_M=\gamma_M z_j$, $j=1,\dots,N-1$
can be obtained from the evaluation at $w_M=\gamma_M z_N$
using Property (2).
This idea is essentially the same with the Izergin-Korepin analysis
for the domain wall boundary partition functions.

For the case of projected wavefunctions,
there is another case we have to consider: the case when $x_N \neq M$.
For this case, it is easier to connect the projected wavefunctions
from its graphical description, and we have
\eqref{ordinaryrecursionwavefunction2}.
Note that the smaller projected wavefunctions connected is \\
$W_{M-1,N}(z_1,\dots,z_{N}|w_1,\dots,w_{M-1}|x_1,\dots,x_{N})$
which is different from the one for the case when $x_N=M$.

In both cases $x_N=M$ and $x_N \neq M$,
we are able to connect the projected wavefunctions of different sizes,
and continuing this process successively,
the relations can be regarded as recursion relations between
projected wavefunctions.
For the Izergin-Korepin analysis to be successful
such that it gives the uniqueness of the projected wavefunctions,
we need the intitial condition for the recursion relations,
and it is Property (4) in
Proposition \ref{ordinarypropertiesfordomainwallboundarypartitionfunction}.
Hence, if one finds an explicit function
satisfying all the properties in
Proposition \ref{ordinarypropertiesfordomainwallboundarypartitionfunction},
it is the one representing the projected wavefunctions.
This is given in the next section.

\section{Generalized Schur functions}

\begin{definition}
We define the following symmetric function \\
$F_{M,N}(z_1,\dots,z_N|w_1,\dots,w_M|x_1,\dots,x_N)$
which depends on the symmetric variables $z_1,\dots,z_N$,
complex parameters $w_1,\dots,w_M$, $\alpha_1,\dots,\alpha_M$,
$\gamma_1,\dots,\gamma_M$
and integers $x_1,\dots,x_N$ satisfying
$1 \le x_1 < \cdots < x_N \le M$,
\begin{align}
&F_{M,N}(z_1,\dots,z_N|w_1,\dots,w_M|x_1,\dots,x_N) \nonumber \\
=
&\prod_{j=1}^N w_{x_j} \frac{1}{\prod_{1 \le j < k \le N}(z_k-z_j)}
\sum_{\sigma \in S_N} (-1)^\sigma
\prod_{j=1}^N \prod_{k=x_j+1}^M (w_k-\gamma_k z_{\sigma(j)})
\nonumber \\
&\times
\prod_{j=1}^N \prod_{k=1}^{x_j-1}\{(1-\alpha_k \gamma_k) z_{\sigma(j)}
+\alpha_k w_k \}.
\label{ordinaryrighthandside}
\end{align}
\end{definition}

The symmetric function
$F_{M,N}(z_1,\dots,z_N|w_1,\dots,w_M|x_1,\dots,x_N)$
\eqref{ordinaryrighthandside}
is a generalization of the (factorial) Schur functions.
\eqref{ordinaryrighthandside} can be rewritten in the form
using Young diagrams as
\begin{align}
F_{M,N}(z_1,\dots,z_N|w_1,\dots,w_M|x_1,\dots,x_N)
=\frac{F_{\lambda+\delta}({\bf z}|w_1,\dots,w_M)}
{\prod_{1 \le j < k \le N}(z_j-z_k)}. \label{correspondence}
\end{align}
Here, ${\bf z}=\{z_1,\dots,z_N \}$ is a set of variables
and $\lambda$ denotes a Young diagram
$\lambda=(\lambda_1,\lambda_2,\dots,\lambda_N)$
with weakly decreasing non-negative integers
$\lambda_1 \ge \lambda_2 \ge \cdots \ge \lambda_N \ge 0$,
and $\delta=(N-1,N-2,\dots,0)$.
$F_{\mu}({\bf z}|\{ \alpha \}|\{ \gamma \})$
is an $N \times N$ determinant
\begin{align}
F_{\mu}({\bf z}|w_1,\dots,w_M)
=\mathrm{det}_N
(
f_{\mu_j}(z_k|w_1,\dots,w_M)
),
\end{align}
where
\begin{align}
f_\mu(z|w_1,\dots,w_M)
=w_{\mu+1}
\prod_{j=1}^\mu \{ (1-\alpha_j \gamma_j)z+\alpha_j w_j \}
\prod_{j=\mu+2}^M(w_j-\gamma_j z).
\end{align}
The Young diagrams
$\lambda=(\lambda_1,\lambda_2,\dots,\lambda_N) \in \mathbb{Z}^N$
($M-N \ge \lambda_1 \ge \lambda_2 \ge \dots \ge \lambda_N \ge 0$)
in the form \eqref{correspondence}
and the sequence of integers $x_1,\dots,x_N$ satisfying
$1 \le x_1 < \cdots < x_N \le M$
in \eqref{ordinaryrighthandside} representing
$F_{M,N}(z_1,\dots,z_N|w_1,\dots,w_M|x_1,\dots,x_N)$
are connected by the translation rule
$\lambda_j=x_{N-j+1}-N+j-1$, $j=1,\dots,N$.

In the limit $w_j=1$, $j=1,\dots,M$, $\gamma_j=0$, $j=1,\dots,M$,
we can see from the form \eqref{correspondence} that
$F_{M,N}(z_1,\dots,z_N|w_1,\dots,w_M|x_1,\dots,x_N)$
reduces to the factorial Schur functions.
If one furthermore sets $\alpha_j=0$, $j=1,\dots,M$,
it further reduces to the Schur functions.

We have the following correspondence between the
projected wavefunctions of the integrable model
and the generalized Schur function
$F_{M,N}(z_1,\dots,z_N|w_1,\dots,w_M|x_1,\dots,x_N)$.

\begin{theorem}
The projected wavefunctions of the generalized free-fermion model
\\
$W_{M,N}(z_1,\dots,z_N|w_1,\dots,w_M|x_1,\dots,x_N)$
is explicitly expressed as the
product of factors \\
$\prod_{1 \le j < k \le N}(z_j+tz_k)$ and 
the symmetric function
$F_{M,N}(z_1,\dots,z_N|w_1,\dots,w_M|x_1,\dots,x_N)$
\begin{align}
&W_{M,N}(z_1,\dots,z_N|w_1,\dots,w_M|x_1,\dots,x_N) \nonumber \\
=&
\prod_{1 \le j < k \le N}(z_j+tz_k)
F_{M,N}(z_1,\dots,z_N|w_1,\dots,w_M|x_1,\dots,x_N).
\label{maintheorem}
\end{align}
\end{theorem}

In the limit $w_j=1$, $j=1,\dots,M$, $\gamma_j=0$, $j=1,\dots,M$,
\eqref{maintheorem} reduces to the main theorem
of Bump-McNamara-Nakasuji \cite{BMN}.
Taking the limit $\alpha_j=0$, $j=1,\dots,M$ furthermore,
one gets the main theorem of Bump-Brubaker-Friedberg \cite{BBF}.

\begin{proof}
Let us denote the right hand side of \eqref{maintheorem} as
$G_{M,N}(z_1,\dots,z_N|w_1,\dots,w_M|x_1,\dots,x_N)$.
\begin{align}
&G_{M,N}(z_1,\dots,z_N|w_1,\dots,w_M|x_1,\dots,x_N) \nonumber \\
:=&\prod_{1 \le j < k \le N}(z_j+tz_k)
F_{M,N}(z_1,\dots,z_N|w_1,\dots,w_M|x_1,\dots,x_N)
\nonumber \\
=&\prod_{j=1}^N w_{x_j} \frac{\prod_{1 \le j < k \le N}(z_j+tz_k)}{\prod_{1 \le j < k \le N}(z_k-z_j)}
\sum_{\sigma \in S_N} (-1)^\sigma
\prod_{j=1}^N \prod_{k=x_j+1}^M (w_k-\gamma_k z_{\sigma(j)})
\nonumber \\
&\times
\prod_{j=1}^N \prod_{k=1}^{x_j-1}\{(1-\alpha_k \gamma_k) z_{\sigma(j)}
+\alpha_k w_k \}.
\end{align}

We prove this theorem by showing that
$G_{M,N}(z_1,\dots,z_N|w_1,\dots,w_M|x_1,\dots,x_N)$
satisfies all the four Properties in
Proposition \ref{ordinarypropertiesfordomainwallboundarypartitionfunction}.

To show Property (1),
first note that the factor
$\displaystyle \prod_{j=1}^N \prod_{k=x_j+1}^M (w_k-\gamma_k z_{\sigma(j)})$
in \\
$G_{M,N}(z_1,\dots,z_N|w_1,\dots,w_M|x_1,\dots,x_N)$
is a polynomial of degree $N-1$ in $w_M$ if $x_N=M$
and degree $N$ if $x_N \neq M$.
For the case $x_N \neq M$,
one sees that the dependence on $w_N$ just only comes from this factor.
For the case $x_N=M$, there is a factor $w_M$ coming from
the overall factor $\prod_{j=1}^N w_{x_j}$.
Thus, $G_{M,N}(z_1,\dots,z_N|w_1,\dots,w_M|x_1,\dots,x_N)$
is a polynomial of degree $N$ in $w_M$
for both cases $x_N=M$ and $x_N \neq M$, hence Property (1) is proved.

Let us next show Property (2).
First, note that $F_{M,N}(z_1,\dots,z_N|w_1,\dots,w_M|x_1,\dots,x_N)$
which is a part of $G_{M,N}(z_1,\dots,z_N|w_1,\dots,w_M|x_1,\dots,x_N)$
is symmetric with respect to $z_1,\dots,z_N$ since both the denominator
$\prod_{1 \le j < k \le N}(z_k-z_j)$ and the numerator
\begin{align}
&\prod_{j=1}^N w_{x_j}
\sum_{\sigma \in S_N} (-1)^\sigma
\prod_{j=1}^N \prod_{k=x_j+1}^M (w_k-\gamma_k z_{\sigma(j)})
\prod_{j=1}^N \prod_{k=1}^{x_j-1}\{(1-\alpha_k \gamma_k) z_{\sigma(j)}
+\alpha_k w_k \},
\end{align}
are
antisymmetric with respect to simple permutations of the spectral parameters
$z_1,\dots,z_N$.
This means
\begin{align}
F_{M,N}(z_1,\dots,z_N|w_1,\dots,w_M|x_1,\dots,x_N)
=F_{M,N}(z_{\sigma(1)},\dots,z_{\sigma(N)}|w_1,\dots,w_M|x_1,\dots,x_N).
\label{forpermutationone}
\end{align}
Looking at the other factor $\prod_{1 \le j < k \le N}(z_j+tz_k)$
which constructs the function \\
$G_{M,N}(z_1,\dots,z_N|w_1,\dots,w_M|x_1,\dots,x_N)$,
we have
\begin{align}
&\prod_{\substack{1 \le j < k \le N \\ \sigma(j) > \sigma(k)}}
(z_{\sigma(j)}+t z_{\sigma(k)})
\prod_{1 \le j < k \le N}(z_j+tz_k) \nonumber \\
=&
\prod_{\substack{1 \le j < k \le N \\ \sigma(j) > \sigma(k)}}
(z_{\sigma(k)}+t z_{\sigma(j)})
\prod_{1 \le j < k \le N}(z_{\sigma(j)}+tz_{\sigma(k)}).
\label{forpermutationtwo}
\end{align}
From \eqref{forpermutationone}, \eqref{forpermutationtwo}
and the fact that $G_{M,N}(z_1,\dots,z_N|w_1,\dots,w_M|x_1,\dots,x_N)$
is defined as a product of $\prod_{1 \le j < k \le N}(z_j+tz_k)$
and $F_{M,N}(z_1,\dots,z_N|w_1,\dots,w_M|x_1,\dots,x_N)$,
we have
\begin{align}
&\prod_{\substack{1 \le j < k \le N \\ \sigma(j) > \sigma(k)}}
(z_{\sigma(j)}+t z_{\sigma(k)})
G_{M,N}(z_1,\dots,z_N|w_1,\dots,w_M|x_1,\dots,x_N) \nonumber \\
=&
\prod_{\substack{1 \le j < k \le N \\ \sigma(j) > \sigma(k)}}
(z_{\sigma(k)}+t z_{\sigma(j)})
G_{M,N}(z_{\sigma(1)},\dots,z_{\sigma(N)}|w_1,\dots,w_M|x_1,\dots,x_N).
\end{align}
We have shown that
$G_{M,N}(z_1,\dots,z_N|w_1,\dots,w_M|x_1,\dots,x_N)$
satisfies the same relation
the projected wavefucntions 
$W_{M,N}(z_1,\dots,z_N|w_1,\dots,w_M|x_1,\dots,x_N)$
must satisfy. Hence Property (2) is proved.

Next we show Property (3).
We first prove that the function \\
$G_{M,N}(z_1,\dots,z_N|w_1,\dots,w_M|x_1,\dots,x_N)$ satisfies
\eqref{ordinaryrecursionwavefunction}
and \eqref{ordinaryrecursionwavefunctionaddition}
for the case $x_N=M$.
To prove \eqref{ordinaryrecursionwavefunction},
we first note that the factor
\begin{align}
\prod_{j=1}^N \prod_{k=x_j+1}^M (w_k-\gamma_k z_{\sigma(j)}),
\end{align}
in each summand essentially becomes
\begin{align}
\prod_{j=1}^{N-1} \prod_{k=x_j+1}^M (w_k-\gamma_k z_{\sigma(j)}).
\label{ordinaryfactorconsideration}
\end{align}
Concentrating on the factor
$\displaystyle
\prod_{j=1}^{N-1} (w_M-\gamma_M z_{\sigma(j)})
$ from \eqref{ordinaryfactorconsideration},
one finds this factor vanishes unless $\sigma$ satisfies $\sigma(N)=N$
if one substitutes $w_M=\gamma_M z_N$.

Therefore, only the summands satisfying $\sigma(N)=N$ 
in \eqref{ordinaryrighthandside} survive
after the substitution $w_M=\gamma_M z_N$.
Keeping this in mind, one rewrites
$G_{M,N}(z_1,\dots,z_N|w_1,\dots,w_M|x_1,\dots,x_N)$
evaluated at $w_M=\gamma_M z_N$
by using the symmetric group $S_{N-1}$
where every $\sigma^\prime \in S_{N-1}$ satisfies
$\{\sigma^\prime(1),\cdots,\sigma^\prime(N-1)\}=\{1,\cdots,N-1 \}$ as follows:
\begin{align}
&G_{M,N}(z_1,\dots,z_N|w_1,\dots,w_M|x_1,\dots,x_N)
|_{w_M=\gamma_M z_N} \nonumber \\
=&\gamma_M z_N \prod_{j=1}^{N-1} w_{x_j}
\frac{\prod_{1 \le j < k \le N-1}(z_j+tz_k)
\prod_{j=1}^{N-1}(z_j+tz_N)}{\prod_{1 \le j < k \le N-1}(z_k-z_j)
\prod_{j=1}^{N-1}(z_N-z_j)} \nonumber \\
\times&\sum_{\sigma^\prime \in S_{N-1}}
(-1)^{\sigma^\prime}
\prod_{j=1}^{N-1} \prod_{k=x_j+1}^{M-1}(w_k-\gamma_k z_{\sigma^\prime(j)})
\prod_{j=1}^{N-1}\gamma_M(z_N-z_{\sigma^\prime(j)})
\nonumber \\
\times&
\prod_{j=1}^{N-1} \prod_{k=1}^{x_j-1}
\{(1-\alpha_k \gamma_k)z_{\sigma^\prime(j)}+\alpha_k w_k \}
\prod_{k=1}^{M-1}
\{(1-\alpha_k \gamma_k)z_N+\alpha_k w_k \}.
\label{ordinaryrighthandsideaftersubstitution}
\end{align}
One easily notes that
the factors $\displaystyle \prod_{k=1}^{M-1}
\{(1-\alpha_k \gamma_k)z_N+\alpha_k w_k \}
$ in the sum are independent of the permutation
$S^\prime_{N-1}$.
One also finds that the product of factors
$\displaystyle \frac{1}{\prod_{j=1}^{N-1}(z_N-z_j)}$
and
$\displaystyle \prod_{j=1}^{N-1}\gamma_M(z_N-z_{\sigma^\prime(j)})$
can be simplified as
\begin{align}
&\frac{1}{\prod_{j=1}^{N-1}(z_N-z_j)}
\prod_{j=1}^{N-1}\gamma_M(z_N-z_{\sigma^\prime(j)})
=\frac{1}{\prod_{j=1}^{N-1}(z_N-z_j)}
\prod_{j=1}^{N-1}\gamma_M(z_N-z_j)
=\gamma_M^{N-1}.
\end{align}
Thus, \eqref{ordinaryrighthandsideaftersubstitution}
can be rewritten furthermore as
\begin{align}
&G_{M,N}(z_1,\dots,z_N|w_1,\dots,w_M|x_1,\dots,x_N)
|_{w_M=\gamma_M z_N} \nonumber \\
=&\gamma_M^N z_N
\prod_{j=1}^{N-1}(z_j+tz_N)
\prod_{j=1}^{M-1}
\{(1-\alpha_j \gamma_j)z_N+\alpha_j w_j \}
\nonumber \\
\times&
\prod_{j=1}^{N-1} w_{x_j}
\frac{\prod_{1 \le j < k \le N-1}(z_j+tz_k)}
{\prod_{1 \le j < k \le N-1}(z_k-z_j)}
\sum_{\sigma^\prime \in S_{N-1}}
(-1)^{\sigma^\prime}
\prod_{j=1}^{N-1} \prod_{k=x_j+1}^{M-1}(w_k-\gamma_k z_{\sigma^\prime(j)})
\nonumber \\
\times&
\prod_{j=1}^{N-1} \prod_{k=1}^{x_j-1}
\{(1-\alpha_k \gamma_k)z_{\sigma^\prime(j)}+\alpha_k w_k \}.
\label{ordinaryrighthandsideaftersubstitutiontwo}
\end{align}
Since
\begin{align}
&\prod_{j=1}^{N-1} w_{x_j}
\frac{\prod_{1 \le j < k \le N-1}(z_j+tz_k)}
{\prod_{1 \le j < k \le N-1}(z_k-z_j)}
\sum_{\sigma^\prime \in S_{N-1}}
(-1)^{\sigma^\prime}
\prod_{j=1}^{N-1} \prod_{k=x_j+1}^{M-1}(w_k-\gamma_k z_{\sigma^\prime(j)})
\nonumber \\
\times&
\prod_{j=1}^{N-1} \prod_{k=1}^{x_j-1}
\{(1-\alpha_k \gamma_k)z_{\sigma^\prime(j)}+\alpha_k w_k \} \nonumber \\
=&
G_{M-1,N-1}(z_1,\dots,z_{N-1}|w_1,\dots,w_{M-1}|x_1,\dots,x_{N-1})
,
\end{align}
one finds that
\eqref{ordinaryrighthandsideaftersubstitutiontwo} is nothing but the
following recursion relation
for the function \\
$G_{M,N}(z_1,\dots,z_N|w_1,\dots,w_M|x_1,\dots,x_N)$
\begin{align}
&G_{M,N}(z_1,\dots,z_N|w_1,\dots,w_M|x_1,\dots,x_N)
|_{w_M=\gamma_M z_N} \nonumber \\
=&\gamma_M^N z_N
\prod_{j=1}^{N-1}(z_j+tz_N)
\prod_{j=1}^{M-1}
\{(1-\alpha_j \gamma_j)z_N+\alpha_j w_j \}
\nonumber \\
\times&G_{M-1,N-1}(z_1,\dots,z_{N-1}|w_1,\dots,w_{M-1}|x_1,\dots,x_{N-1}),
\end{align}
which is exactly the same recursion relation the projected wavefunctions
\\
$W_{M,N}(z_1,\dots,z_N|w_1,\dots,w_M|x_1,\dots,x_N)$ must satisfy,
hence \eqref{ordinaryrecursionwavefunction} is shown.
\eqref{ordinaryrecursionwavefunctionaddition} can be shown immediately
since $G_{M,N}(z_1,\dots,z_N|w_1,\dots,w_M|x_1,\dots,x_N)$
evaluated at $w_M=0$ becomes zero due to
the overall factor $\prod_{j=1}^N w_{x_j}$
in $F_{M,N}(z_1,\dots,z_N|w_1,\dots,w_M|x_1,\dots,x_N)$
and the fact that we are dealing the case $x_N=M$.

Now let us examine the case $x_N \neq M$.
This can be shown in a similar but much simpler way.
We rewrite $G_{M,N}(z_1,\dots,z_N|w_1,\dots,w_M|x_1,\dots,x_N)$ as
\begin{align}
&G_{M,N}(z_1,\dots,z_N|w_1,\dots,w_M|x_1,\dots,x_N) \nonumber \\
=&\prod_{j=1}^N w_{x_j} \frac{\prod_{1 \le j < k \le N}(z_j+tz_k)}{\prod_{1 \le j < k \le N}(z_k-z_j)}
\sum_{\sigma \in S_N} (-1)^\sigma
\prod_{j=1}^N \prod_{k=x_j+1}^{M-1} (w_k-\gamma_k z_{\sigma(j)})
\prod_{j=1}^N (w_M-\gamma_M z_{\sigma(j)})
\nonumber \\
&\times
\prod_{j=1}^N \prod_{k=1}^{x_j-1}\{(1-\alpha_k \gamma_k) z_{\sigma(j)}
+\alpha_k w_k \}.
\label{originalrighthandsidenew}
\end{align}
Noting
\begin{align}
\displaystyle 
\prod_{j=1}^N (w_M-\gamma_M z_{\sigma(j)})
=\prod_{j=1}^N (w_M-\gamma_M z_j),
\end{align}
we can take this factor out of the sum in
\eqref{originalrighthandsidenew} and we get
\begin{align}
&G_{M,N}(z_1,\dots,z_N|w_1,\dots,w_M|x_1,\dots,x_N)
\nonumber \\
=&\prod_{j=1}^N (w_M-\gamma_M z_j)
\prod_{j=1}^N w_{x_j} \frac{\prod_{1 \le j < k \le N}(z_j+tz_k)}{\prod_{1 \le j < k \le N}(z_k-z_j)}
\sum_{\sigma \in S_N} (-1)^\sigma
\prod_{j=1}^N \prod_{k=x_j+1}^{M-1} (w_k-\gamma_k z_{\sigma(j)})
\nonumber \\
&\times
\prod_{j=1}^N \prod_{k=1}^{x_j-1}\{(1-\alpha_k \gamma_k) z_{\sigma(j)}
+\alpha_k w_k \} \nonumber \\
=&\prod_{j=1}^N (w_M-\gamma_M z_j)
G_{M-1,N}(z_1,\dots,z_N|w_1,\dots,w_{M-1}|x_1,\dots,x_N),
\end{align}
which is also exactly the recursion relation the projected wavefunctions \\
$G_{M,N}(z_1,\dots,z_N|w_1,\dots,w_M|x_1,\dots,x_N)$
must satisfy for the case $x_N \neq M$.

Finally it is trivial to check from its definition that
\begin{align}
&
G_{M,1}(z|w_1,\dots,w_M|M)=w_M \prod_{k=1}^{M-1} \{ (1-\alpha_k \gamma_k)z
+\alpha_k w_k \},
\end{align}
hence Property (4) is shown.

Since we have proved that the function
$G_{M,N}(z_1,\dots,z_N|w_1,\dots,w_M|x_1,\dots,x_N)$
satisfies all the
Properties (1), (2), (3) and (4) in
Proposition \ref{ordinarypropertiesfordomainwallboundarypartitionfunction},
we conclude it is the explicit form of the projected wavefunctions \\
$
W_{M,N}(z_1,\dots,z_N|w_1,\dots,w_M|x_1,\dots,x_N)
=
G_{M,N}(z_1,\dots,z_N|w_1,\dots,w_M|x_1,\dots,x_N)
$.
\end{proof}

\section{Conclusion}
In this paper, we studied the generalized
free-fermion model in an external field.
We applied the Izergin-Korepin analysis on the projected wavefunctions
which is a generalization of the Izergin-Korepin analysis on the
domain wall boundary partition functions,
which was recently done for the case of the $U_q(sl_2)$ six-vertex model
in \cite{Motr}.
We extracted the properties about the degree, symmetry, recursion relations
and initial conditions the projected wavefunctions satisfy.
Next we proved that the product of factors and certain symmetric functions
satisfies all the required properties, hence it represents the
projected wavefunctions.
The result can be regarded as an extension of the Tokuyama formula
for the (factorial) Schur functions by Bump-Brubaker-Friedberg \cite{BBF}
and Bump-McNamara-Nakasuji \cite{BMN}.

The result obtained in this paper can also be proved by using
the arguments initiated in \cite{BBF},
which views the partition functions as functions of
the free parameter in the auxiliary spaces.
The Izergin-Korepin analysis used in this paper views the partition functions
as functions of inhomogeneous parameters in the quantum spaces.
The comparison of the two different ways of arguments seems to be interesting.
We use the result obtained in this paper to study algebraic combinatorial
properties of the generalized Schur functions
in our forthcoming paper \cite{MSW}.
Extending the Izergin-Korepin analysis to other boundary conditions
of the generalized free-fermion model is one of the interesting topics
regarding this paper.
There may be some cases which the Izergin-Korepin analysis is suitable,
and some other cases which the arguments in \cite{BBF} are useful.
We think that developing various techniques are useful for the
study of partition functions of integrable lattice models.

\section*{Acknowledgements}
The author sincerely expresses his gratitude to the referees for
helpful comments and advises for the improvement of the paper.
This work was partially supported by grant-in-Aid
for Scientific Research (C) No. 16K05468.

\end{document}